# About group and system-wide lesions of complex network systems and intersystem interactions

Olexandr Polishchuk

Laboratory of Modeling and Optimization of Complex Systems
Pidstryhach Institute for Applied Problems of Mechanics and Mathematics, National Academy of Sciences of Ukraine, Lviv, Ukraine
od_polishchuk@ukr.net

**Abstract** – *The main types of negative internal and external influences on complex network systems (NS) and intersystem interactions in monoflow partially overlapped multilayer network systems (MLNS) are analyzed. Among such influences, targeted attacks on real complex systems and their non-targeted damages of various nature are primarily singled out. On the basis of structural and flow models of NS and MLNS, sequential and simultaneous local, group and system-wide damages to the structure and operation process of complex network systems and intersystem interactions are considered. The existing structural and functional scenarios of consecutive attacks on the most important components of the system and the need to neutralize the sources of such attacks as one of the means of system protection are analyzed. The peculiarities of application of such scenarios for the protection of NS and MLNS from system-wide non-target lesions are considered. Approaches to evaluation and overcoming the consequences of negative impacts on the structure and operation process of complex network systems and intersystem interactions are proposed.*

**Keywords** – *complex network, network system, intersystem interactions, multilayer network system, robustness, vulnerability, targeted attack, non-target lesion*

## ВСТУП

Кожна природна або створена людиною система є уразливою до багатьох внутрішніх та зовнішніх негативних впливів [1, 2]. Клімат та стан біосфери Землі погіршуються під дією індустріального суспільства [3], фінансові кризи та збройні конфлікти негативно впливають на економіку та соціальні настрої населення [4, 5], поширення неправдивої інформації спотворює суспільну думку громадян [6], терористичні або хакерські атаки [7], стихійні лиха та техногенні катастрофи [8, 9], епідемії небезпечних інфекційних захворювань та комп'ютерні віруси [10, 11] тощо створюють загрози для процесу функціонування багатьох реальних систем [12, 13]. У теорії складних мереж (ТСМ) основна увага дослідників зосереджена на вивченні випадкових негативних впливів та цілеспрямованих атак на структуру системи [14, 15]. Проблемі уразливості процесу функціонування складних мережевих систем (МС) приділяється значно менша увага [16]. Безумовно, що ураження структури негативно впливає на процес функціонування системи, однак збої в її роботі можуть виникати і за неураженої структури [17]. Ще однією особливістю сучасних досліджень стійкості МС до різнорідних негативних впливів є розробка сценаріїв уражень окремих елементів або послідовного ураження групи найважливіших зі структурного погляду вузлів мережі [18, 19], хоча очевидно, що одно-часна атака на таку групу або загальносистемна атака, яка у тій або іншій мірі вражає усі елементи МС, є значно небезпечнішою для будь-якої реальної мережевої системи. Поряд із цілеспрямованими атаками, на систему можуть негативно впливати і інші чинники природного або штучного характеру (поширення інфекційних захворювань, затори у великих містах тощо) [20-22], які за багатьма ознаками можуть бути подібними до таких атак (поширення комп'ютерних вірусів, *Ddos*-атаки і т. ін.). Такі впливи часто є непередбачуваними та не мають ні явного «зловмисника», ні наперед визначеної цілі. Вони можуть бути достатньо масштабними і втрати, які заподіюють системі подібні «нецільові» ураження, часто перевищують наслідки масових цілеспрямованих атак [22, 23]. Чітке розуміння можливих способів ураження системи та їх особливостей безумовно допоможе у розробленні дієвих засобів упередження, захисту та подолання наслідків таких уражень.

Кожна реальна МС взаємодіє з багатьма іншими системами [24], у поєднанні з якими вона утворює надсистемні утворення різних типів. Ураження такої МС може призвести до дестабілізації процесу функціонування усіх пов'язаних із нею систем. Це яскраво показало поширення пандемії Covid-19, яке призвело до суттєвого скорочення пасажирських та вантажних транспортних потоків, погіршення фінансово-економічного стану багатьох країн та породженого ним соціального невдоволення населення, до яких виявились неготовими навіть найрозвиненіші держави світу. Це означає, що дослідження проблеми стійкості міжсистемних взаємодій є не менш важливою, ніж визначення уразливості окремої мережевої системи, яка приймає у них участь. Натепер дослідження стійкості таких взаємодій в основному зосереджені на взаємозалежних (interdependent) багатошарових мережах (БШМ), тобто ієрархічно-мережевих структурах прямого підпорядкування з лінійною моделлю управління [25, 26]. Аналізуючи загрози, які можуть порушити структуру або дестабілізувати процес функціонування реальних МС, та розробляючи відповідні засоби їх захисту, дослідники часто абстрагуються від джерел цілеспрямованих атак та нецільових уражень, які можуть бути і внутрішніми, і зовнішніми по відношенню до системи. Водночас, блокування таких джерел є одним із дієвих засобів захисту як окремої МС, так і процесу міжсистемних взаємодій загалом, адже своєчасна нейтралізація терористичної або хакерської групи чи розроблення вакцин та ліків від небезпечних інфекційних захворювань може запобігти тій шкоді, яку вони можуть заподіяти.

**Мета статті** – дослідження уразливості складних мережевих систем і міжсистемних взаємодій до різнорідних групових та загальносистемних внутрішніх та зовнішніх негативних впливів та розроблення способів попередження, боротьби та подолання наслідків таких впливів. Перша частина статті присвячена огляду та аналізу основних типів уражень мережевих систем та підходів до їх захисту від таких уражень.

# ЦІЛЕСПРЯМОВАНІ АТАКИ ТА НЕЦІЛЬОВІ УРАЖЕННЯ СКЛАДНИХ МЕРЕЖЕВИХ СИСТЕМ

Різні види уражень будь-якої реальної складної МС потребують різних, інколи прямо протилежних дій, спрямованих на її захист. Тому, щоб забезпечити стійкість системи до різнорідних негативних впливів, необхідно дати відповіді принаймні на наступні питання:

1) які негативні внутрішні та зовнішні впливи можуть порушити структуру та процес функціонування системи;

2) яким чином забезпечити захист системи залежно від типу негативного внутрішнього або зовнішнього впливу, а оскільки будь-яка велика складна система зазвичай не може уборонити усі свої елементи, то які з них необхідно захистити насамперед;

3) до яких наслідків, незважаючи на всі задіяні засоби захисту, може призвести ураження системи певного типу та яким чином їх оцінити та подолати.

Під внутрішніми негативними впливами розумітимемо ураження системи, спричинені джерелами, які входять до її складу (корупція в країні, аварії на небезпечних виробництвах і т. ін.). Зовнішні негативні впливи породжуються джерелами, які знаходяться поза межами системи (іноземні терористичні або хакерські групи, зовнішня агресія тощо). У цій статті ми розглядаємо два основні види внутрішніх та зовнішніх негативних впливів – цілеспрямовані атаки та нецільові ураження складних мережевих систем та міжсистемних взаємодій. Відмінною рисою цілеспрямованих атак є їх умисність та штучний характер (хакерські та терористичні напади, військові дії та економічні санкції, поширення неправдивої інформації тощо). Особливістю уражень цього типу є наявність цілі атаки, спрямованої на завдання якнайбільшої матеріальної та/або моральної шкоди атакованій системі, та зловмисника, який цю атаку здійснює. На відміну від цілеспрямованих атак, до нецільових уражень можна віднести різнорідні неумисні негативні впливи природного або штучного характеру, виникнення та наслідки яких людина не може своєчасно передбачити (природні та техногенні катастрофи, поширення небезпечних захворювань і т. ін.). Труднощі точної класифікації типу ураження часто бувають пов'язані з невизначеністю його джерела, адже умисний підпал лісу та необачно кинутий недопалок або диверсія на підприємстві та використання неякісного обладнання часто призводять до схожих наслідків. Водночас подібні причини (епідемії коронавірусів SARS Cov-1 у 2002 році в Китаї, MERS у 2010 році на Близькому Сході та SARS Cov-2 у 2019 році в Ухані) можуть призводити до різних наслідків. Багаточинниковість впливів, наявність багатьох спільних ознак, спосіб дії і часто суб'єктивний фактор, які практично неможливо врахувати або передбачити, є ще однією причиною складності класифікації системних уражень. Слід також враховувати, що негативний з боку суб'єкта може бути позитивним з боку об'єк-та впливом (впровадження нових технологій, які витісняють застарілі, санкції проти країн-агресорів тощо).

Ураження системи можуть бути очікуваними (умовно передбачуваними) та неочікуваними. Так, очікуваними можна назвати землетруси у регіонах, які знаходяться на стику тектонічних плит (Каліфорнія, Чилі, Японія), потужні урагани у Карибському басейні та Південно-східній Азії, лісові пожежі на сході Австралії та у Каліфорнії, епідемії лихоманки Ебола в Центральній Африці тощо. До неочікуваних можна віднести глобальне поширення СНІДу та Covid-19 або землетрус в Гаїті силою 7 балів за шкалою Ріхтера у 2010 році, жертвами якого стали більше 200 тис. чоловік та понад мільйон людей залишились без притулку. Водночас жертвами землетрусу магнітудою у 8.8 балів у Чилі в тому ж році стало менше однієї тисячі чоловік. Очікуваність унаслідок повторюваності землетрусів змусило уряд цієї та інших країн особливо ретельно ставитись до вимог міцності будівель та об'єктів критичної інфраструктури і ефективності роботи відповідних підрозділів надзвичайних ситуацій. Це означає, що підготовлена до захисту від певного ураження система зазнає значно менших втрат, ніж непідготовлена. Однак, попереджувальних заходів може виявитися недостатньо для захисту системи навіть для умовно передбачуваних уражень. Так, після загалом неочікуваної Чорнобильської катастрофи вимоги до безпеки АЕС були значно посилені практично для всіх атомних електростанцій світу. Однак, вони не врахували можливість затоплення охолоджуючих пристроїв унаслідок цунамі, як це сталося у 2011 р. на АЕС у Фукусімі (Японія). Це означає, що навіть системи, які вважаються добре захищеними, можуть бути захищені недостатньо.

Як цілеспрямовані атаки, так і нецільові ураження системи можуть бути локальними, груповими або загальносистемними як з погляду суб'єкта, так і об'єкта ураження, послідовними або одночасними, поширюватися як у просторі, так і у часі, негативно впливаючи на усі пов'язані з ураженими складові МС. Так, терористична група може взяти в заручники школу (м. Беслан), *DDoS*-атаки зазвичай здійснюються з кількох джерел на одну або кілька комп'ютерних мереж, індустріальне суспільство негативно впливає на всю біосферу Землі, а пандемія Covid-19 поширилась усіма країнами світу. У статті [15] було показано, що ураження лише 1% вузлів-доменів Інтернету з найбільшими ступенями удвічі знижує його продуктивність, а блокування 4% таких вузлів поділяє його на незв'язні складові. В Україні кількість державних банків у банківській системі країни на початку 2022 р. не перевищувала 0.7%. Водночас частка їх активів у цій системі була рівною 55.2%, а частка депозитів фізичних осіб – 61.6% [27]. Успішна атака саме на цю, невелику за кількістю, групу банків призведе до найбільших втрат у фінансовій системі держави. Масові *DDoS*-атаки 14 січня та 14-16 лютого 2022 р. на більш ніж 70 найважливіших державних, безпекових, фінансових та соціальних комп'ютерних мереж України [28] можна вважати спробою загальносистемного ура-ження системи її державного управління. Це означає, що для критичної дестабілізації або припинення роботи реальної МС потрібно одночасно заблокувати функціонування певної групи вузлів. Дійсно, послідовні атаки на окремі, навіть найважливіші вузли мережевої системи, як це пропонується у розроблених натепер сценаріях цілеспрямованих атак [12, 18] часто дозволяють перерозподілити їх функції між іншими вузлами. Однак, протидіяти одночасній успішній атаці на групу найважливіших за певними ознаками елементів МС, а головне подолати наслідки такої атаки, значно складніше. Як цілеспрямовані атаки, так і нецільові ураження можуть бути централізованими, коли негативний вплив поширюється від одного джерела, так і децентралізованими, коли уражений елемент сам стає джерелом такого впливу [11], тобто, одиничне ураження може поступово перерости в групове або загальносистемне.

Ураження системи можуть поширюватися з різною швидкістю – від «майже миттєвих» (каскадні явища в електромережах [29]) до тих, які діють десятиліттями (вплив промислового та аграрного виробництва на клімат Землі [2]). Зараження комп'ютерними вірусами займає кілька секунд або хвилин [11], а інкубаційний період інфекційних захворювань – від кількох годин до кількох тижнів [10], дія економічних санкцій – від кількох місяців до кількох років. Очевидно, що різна швидкість ураження вимагає різних за швидкодією засобів захисту або протидії йому. Інколи такі засоби із-за багатьох об'єктивних та суб'єктивних причин неможливо розробити або своєчасно задіяти (вакцини від лихоманки Ебола або Covid-19).

## УРАЖЕННЯ СТРУКТУРИ ТА ПРОЦЕСУ ФУНКЦІОНУВАННЯ СКЛАДНИХ МЕРЕЖЕВИХ СИСТЕМ ТА МІЖСИСТЕМНИХ ВЗАЄМОДІЙ

Вивчаючи реальні мережеві системи та міжсистемні взаємодії різних типів, ми насправді досліджуємо побудовані на підставі емпіричних та теоретичних даних моделі таких систем і взаємодій (структурні, функціональні, інформаційні, математичні тощо). Ідентифікуючи елементи системи, які потребують першочергового захисту, ми визначаємо їх важливість у тій або іншій моделі, базуючись на тих характеристиках елемента, які ця модель дозволяє визначити. Так, структурна модель МС, яка зазвичай відображається у виді $G=(V,E)$, де $V$ – множина вузлів та $E$ – множина поєднуючих ці вузли ребер мережі, повністю описується матрицею суміжності $\mathbf{A}=\{a_{ij}\}_{i,j=1}^{N}$, у якій значення $a_{ij}=1$, якщо існує ребро, яке з'єднує вузли $n_i$ та $n_j$, та $a_{ij}=0$, $i,j=\overline{1,N}$, якщо такого ребра немає, $N$ – кількість вузлів мережі. На основі цієї моделі ми можемо визначати такі структурні показники важливості вузлів, як центральність за ступенем, посередництвом, близькістю, власним значенням [30] і т. ін.

Під динамікою складних мереж у ТСМ зазвичай розуміють зміну складу їх вузлів та структури взаємозв'язків [31]. У той же час кожна реальна система є динамічним утворенням навіть за незмінної структури. Цю динаміку, яка принаймні частково описує процес функціонування системи, можна кількісно відобразити зміною об'ємів потоків, які рухаються мережею. Це природно пояснюється тим, що в одних випадках забезпечення руху потоків є основною ціллю утворення та функціонування таких систем (транспортні, фінансові, торгівельні, інформаційні, соціальні МС і т. ін.), а у інших – процесом, який забезпечує їх життєдіяльність (рух крові, лімфи, нейроімпульсів у людському тілі тощо). Зупинка руху потоків може призвести до суттєвої дестабілізації або навіть припинення існування системи [32]. Розглянемо випадок, коли потоки неперервно розподілені ребрами мережі. Введемо матрицю суміжності $\mathbf{V}(t)$ мережевої системи, структура якої описується матрицею $\mathbf{A}$ [33]. Елементи матриці $\mathbf{V}(t)$ визначаються об'ємами потоків, які пройшли ребрами мережі за період $[t-T, t]$ до поточного моменту часу $t \geq T$, де $T$ – заданий проміжок часу:

$$\mathbf{V}(t) = \{V_{ij}(t)\}_{i,j=1}^{N}, \quad V_{ij}(t) = \frac{\widetilde{V}_{ij}(t)}{\max_{m,l=\overline{1,N}}\{\widetilde{V}_{ml}(t)\}},$$

де

$$\widetilde{V}_{ij}(t) = \int_{t-T}^{t} v_{ij}(\tau)d\tau, \quad t \geq T > 0; \quad v_{ij}(t) = \int_{(n_i, n_j)} \rho_{ij}(t, \mathbf{x})dl, \quad t > 0; \quad \boldsymbol{\rho}(t, x) = \{\rho_{ij}(t, \mathbf{x})\}_{i,j=1}^{N},$$

і $\rho_{ij}(t, \mathbf{x})$ – щільність потоку, який пересувається ребром $(n_i, n_j)$ мережі в поточний момент часу $t$, $\mathbf{x} \in (n_i, n_j) \subset R^m$, $m = 2, 3, ...$, $i, j = \overline{1, N}$, $t \geq T$. Елементи потокової матриці суміжності МС визначаються на основі емпіричних даних про рух потоків її ребрами. Натепер, за допомогою сучасних засобів відбору інформації, такі дані достатньо легко отримати для багатьох природних та переважної більшості створених людиною систем (транспортних, економічних, фінансових, іформаційних і т. ін.) [34]. На основі цієї моделі ми можемо визначити такі показники функціональної важливості елементів МС, як параметри впливу її вузлів, які є глобальним потоковим аналогом структурної центральності за ступенем та визначають важливість вузла в системі як генератора або приймача потоків, та параметри посередництва елементів системи, які є функціональним аналогом структурної центральності посередництва та визначають важливість вузла у МС як транзитера потоків і т. ін. [17].

Більша адекватність функціональних показників порівняно із структурними слідує з наступного прикладу. На рис. 1а відображена бінарна мережа, яка описує структуру фрагменту лише одного шару загальної транспортної системи України, а саме залізничної транспортної системи (ЗТС) її західного регіону. На рис. 1б відображена зважена мережа, яка на основі зафіксованих у певний момент часу значень відповідної потокової матриці суміжності, описує процес функціонування цієї частини ЗТС зі схематичним відображенням об'ємів руху вантажних транспортних потоків, які пройшли ребрами цієї мережі протягом 2020 р. (товщина лі-ній пропорційна вагам – об'ємам потоків). Очевидно, що з функціонального погляду вузол **N₁** зі ступенем 3 є важливішим у системі, ніж вузли **N₂** та **N₃** зі ступенями 4 та 5 відповідно, оскільки через нього проходять значно більші об'єми вантажних потоків. Аналогічно, підмережі **A₁** і **A₂** та **B₁** і **B₂** (рис. 1б та 1в) є абсолютно рівноцінними зі структурного погляду, але відіграють суттєво різну роль у процесі функціонування залізничної транспортної системи регіону [35]. Це означає, що вибір моделі МС чи БШМС та обчислення на її основі показників важливості елементів, може суттєво вплинути на вибір стратегій захисту системи та ефективність цих стратегій.

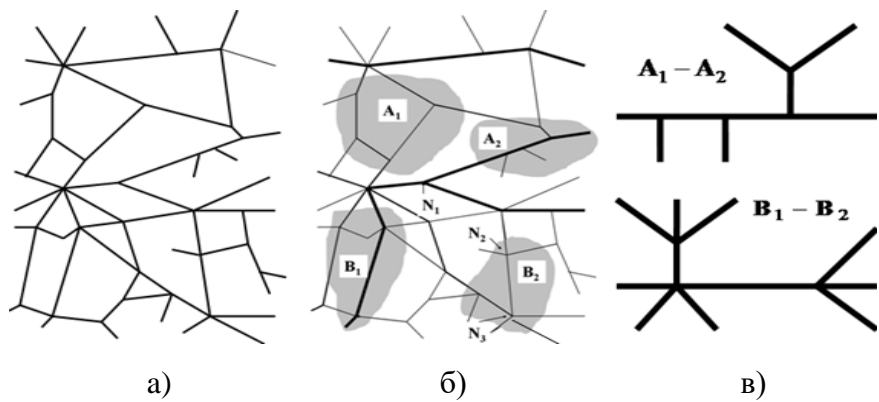

а)                                   б)                                   в)

Рис. 1. Приклади мереж, які відображають структуру та процес функціонування залізничної транспортної системи західного регіону України

Зазвичай ураження елементів структури, наприклад, затор або аварія на міському перехресті, призводить до перерозподілу руху потоків іншими шляхами, що може вже на них створити умови критичного завантаження і, унаслідок цього – затори та блокування нових елементів структури (інших перехресть), як це часто трапляється на автошляхах великого міста у години пік. При цьому, чим більше таких елементів уражено, тим складніше здійснити перерозподіл руху потоків. Тобто, структурні ураження часто породжують збої у процесі функціонування системи і навпаки. Такі процеси часто є взаємопов'язані та можуть мати ланцюговий характер і ураження одного елемента призводити до ураження всієї системи або значної її частини, як це трапляється під час каскадних явищ [36, 37].

Для розроблення способів захисту структури міжсистемних взаємодій необхідне розуміння особливостей цієї структури. Структурна модель таких взаємодій описується багатошаровими мережами (БШМ) [31] та відображається у вигляді

$$G^M = \left( \bigcup_{m=1}^{M} G_m, \bigcup_{\substack{m,k=1 \\ m \neq k}}^{M} E_{mk} \right), \tag{1}$$

де $G_m = (V_m, E_m)$ визначає структуру $m$-го мережевого шару БШМ; $V_m$ – множина вузлів мережі $G_m$; $E_m$ – множина зв'язків мережі $G_m$; $E_{mk}$ – множина зв'язків між вузлами множин $V_m$ та $V_k$, $m \neq k$, $m, k = \overline{1, M}$, де $M$ – кількість шарів БШМ (мережевих систем, які приймають участь у міжсистемних взаємодіях). Множину

$$V^M = \bigcup_{m=1}^{M} V_m$$

називатимемо загальною сукупністю вузлів БШМ, $N^M$ – кількість елементів $V^M$.

Багатошарова мережа $G^M$ повністю описується матрицею суміжності $\mathbf{A}^M = \{\mathbf{A}^{km}\}_{m,k=1}^{M}$, у якій значення $a_{ij}^{km} = 1$, якщо існує ребро, яке з'єднує вузли $n_i^k$ та $n_j^m$, та $a_{ij}^{km} = 0$, $i, j = \overline{1, N^M}$, якщо такого ребра немає. При цьому блоки $\mathbf{A}^{mm}$ описують структуру внутрішньошарових, а блоки $\mathbf{A}^{km}$, $m \neq k$, $m, k = \overline{1, M}$, – міжшарових взаємодій. Якщо усі блоки матриці $\mathbf{A}^M$ визначаються для загальної сукупності вузлів БШМ, то знімається проблема координації номерів вузлів у випадку їх незалежної нумерації для кожного шару. Зі структурного погляду найбільш загальним видом багатошарових мереж можна вважати частково покриті БШМ, перетин множин вузлів $V_m$ яких є непорожнім (рис. 2) [38].

Граничними випадками частково покритих багатошарових мереж є мультиплекси, тобто БШМ, множини вузлів $V_m$ яких повністю співпадають [39], та мультимережі, тобто БШМ, множини вузлів $V_m$ яких не перетинаються [40], $m = \overline{1, M}$. Як і у випадку звичайних МС, структурна модель багатошарової мережевої системи (БШМС) та обчислені на її основі локальні та глобальні характеристики елементів багатошарової мережі (центральності різних типів) дозволяють визначати найважливіші зі структурного погляду її складові [32].

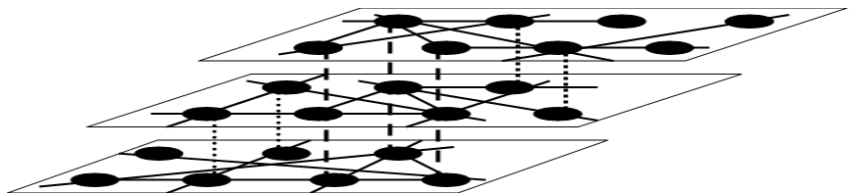

Рис. 2. Приклад структури частково покритої багатошарової мережі

Міжсистемні взаємодії можна описати у вигляді потокової моделі багатошарової мережевої системи, кожний шар якої відображає процес функціонування окремої МС, яка приймає участь у таких взаємодіях [41]. Найбільш загальним типом багатошарових систем є багатовимірні (багатофункціональні, мультипотокові [42]) БШМС, кожний шар яких забезпечує рух специфічного, тобто відмінного від інших шарів, типу потоку. Прикладами таких БШМС є система міжнародного співробітництва, яке включає в себе фінансовий, економічний, військовий, безпековий, науковий, культурний, спортивний та інші шари; система забезпечення життєдіяльності великого міста або регіону країни, до складу якої входять системи постачання електроенергії, газу, води, телефонного та Інтернет зв'язку, кабельного телебачення та охорони і т. ін. Багатовимірні БШМС загалом характеризуються неможливістю переходу потоку з одного шару на інший (перетворення пасажирів у вантажі або команди спортсменів-важкоатлетів у балетну трупу). Тому у [33] для спрощення дослідження міжсистемних взаємодій був запропонований підхід, який полягає у декомпозиції багатовимірних БШМС на монопотокові багатошарові системи, які забезпечують рух потоків лише одного типу різними носіями або системами-операторами та можливість переходу потоку з одного шару на інший через так звані точки переходу, тобто вузли, які мають зв'язки з елементами інших шарів-систем. Точки переходу можуть мати множинні зв'язки з вузлами інших шарів (ефектронна пошта або телефонний зв'язок) або забезпечувати зв'язок лише з точками із тим же номером із загальної сукупності вузлів багатошарової мережі (загальна транспортна БШМС). З функціонального погляду останній випадок означає, що відповідний вузол є елементом кількох систем-шарів та виконує у них одну функцію різними способами. За допомогою цього підходу загальну транспортну систему доцільно поділяти на дві монопотокові чотиришарові системи, кожна з яких забезпечує рух пасажирських або вантажних потоків автомобільним, залізничним, авіаційним або водним (морським чи річковим) транспортом відповідно. Монопотокові БШМС є достатньо розповсюдженими у реальному світі та міжшарові взаємодії у них зазвичай є найбільш інтенсивними. З одного боку такі БШМС є менш уразливими, ніж окремі її шари, унаслідок можливості «дублювання» шляхів руху потоків , а з іншого – саме у таких БШМС ураження одного шару може найбільше вплинути на процес функціонування інших шарів-систем. У них також пришвидшується поширення негативних впливів різної природи (епідемії, комп'ютерні віруси і т. ін.). Процес внутрішньо та міжсистемних взаємодій у монопотокових БШМС можна кількісно описати за допомогою потокової матриці суміжності $\mathbf{V}^M(t)$ [33], структура якої співпадає зі структурою матриці $\mathbf{A}^M$. Блоки $\mathbf{V}^{mm}(t)$ цієї матриці є потоковими матрицями суміжності $m$-го шару-системи БШМС, а блоки $\mathbf{V}^{km}(t)$, $k \neq m$, $k, m = \overline{1, M}$, – потоковими матрицями міжшарових взаємодій у точках переходу багатошарової мережевої системи. На основі цієї моделі ми можемо визначити такі показники функціональної важливості, як потокові ступені та параметри впливу вузлів і параметри посередництва вузлів та ребер БШМС і т. ін. [17].

Очевидно, що проблема уразливості багатошарової мережевої системи є набагато глибшою та складнішою, ніж проблема стійкості до негативних впливів окремого її шару. Ми ви-діляємо два основні типи уражень БШМС. Перший із них полягає у початковому ураженні одного шару, яке тим або іншим чином послідовно призводить до ураження структури або процесу функціонування інших шарів-систем, тобто, уражений шар-система стає зовнішнім джерелом негативного впливу на пов'язані із ним системи. Добре відомі ураження мереж електропостачання великих міст (Нью-Йорк, 1977 р.) або регіонів країни (США та Італія, 2003 р.), які стали причиною збоїв у роботі багатьох промислових систем та систем життєзабезпечення цього міста або регіону. Пандемія Covid-19 призвела до дестабілізації роботи практично всіх економічних, фінансових, транспортних та соціальних систем більшості країн світу. Блокування під час російсько-української війни морських портів України (водного шару загальної транспортної системи країни), через які здійснювалось близько 60% її експорту, не лише створило загрозу голоду 400 млн. чоловік у 38 країнах світу, але й приблизно на 40% скоротило об'єми залізничних та автомобільних вантажних транспортних потоків усередині країни, які забезпечували цей експорт. Другий спосіб ураження БШМС полягає в одночасних негативних впливах на частину або усі її шари. До прикладів таких впливів належать гібридні війни, які здійснюються шляхом поєднання економічних, фінансових, інформаційних, військових уражень або «всеосяжні санкції» проти країн, які несуть загрозу світовій безпеці, тощо. Саме багатовимірні БШМС дозволяють достатньо адекватно моделювати такі впливи та їх наслідки.

## ОСНОВНІ ПІДХОДИ ДО ЗАХИСТУ ВІД ЦІЛЕСПРЯМОВАНИХ АТАК ТА НЕЦІЛЬОВИХ УРАЖЕНЬ

Оскільки ураження БШМС зазвичай здійснюється шляхом послідовного або одночасного ураження окремих її шарів, то підходи до захисту міжсистемних взаємодій формуються насамперед на підставі розроблення способів захисту МС, які входять до складу багатошарової мережевої системи. Першим кроком під час формування таких способів є побудова так званих сценаріїв цілеспрямованих атак на МС [16, 43], тобто послідовності дій, за допомогою яких потенційний «зловмисник» намагатиметься заподіяти якомога більших матеріальних та/або моральних втрат атакованій системі. Тому найбільш дієві сценарії атак формуються, коли їх розробник «ставить себе» на місце зловмисника, який мінімальними засобами намагається завдати максимальної шкоди найважливішим зі структурного та/або функціонального погляду елементам системи. Розробці кожного сценарію повинно передувати вироблення критеріїв успішності атаки. Зі структурного погляду такими критеріями можуть бути поділ МС на незв'язні складові, ліквідація міжсистемних взаємодій між окремими шарами БШМ і т. ін. [13, 19], а з функціонального – припинення або суттєве обмеження руху потоків чи створення умов критичного навантаження найважливіших елементів чи підмереж МС або БШМС [16, 44].

Розроблені натепер структурні сценарії ураження МС, які можна поділити на дві основні групи, базуються на використанні різних показників важливості вузлів у структурі системи (центральностей за ступенем, посередництвом, близькістю, власним значенням тощо) [30]. Кожний із сценаріїв першої групи починається із впорядкування множини вузлів МС у порядку зменшення значень їх центральності відповідного типу та подальшому послідовному вилученні зі структури вузлів згідно цього порядку. Сценарії цієї групи не передбачають зміну значень центральностей вузлів, які залишились у мережі. У другій групі сценаріїв враховується, що з кожним вилученням вузла структура МС нехай незначно, але може змінитися завдяки налагодженню нових зв'язків між вузлами, що залишились. Це потребує нового впорядкування послідовності вузлів МС згідно змінених значень їх центральностей. На наступному кроці цієї групи сценаріїв вилучається вузол з початку новоствореного списку, який врахує ці зміни. Існування більш ніж десяти видів структурних центральностей означає певну неоднозначність їх використання у якості показників важливості вузлів у МС, що було підтверджено на прикладі достатньо простої мережі у [45].

Побудова функціональних сценаріїв цілеспрямованих атак здійснюється за тими ж принципами, що й структурних, з тією різницею, що у якості показників важливості вузлів МС використовуються більш адекватні, як це було показано вище, параметри їх потокового впливу та посередництва [17], що суттєво підвищує їх дієвість. Розглянуті вище структурний та функціональний підходи до побудови сценаріїв цілеспрямованих атак на систему можна поєднати. Наприклад, якщо у послідовності вузлів МС існують групи з однаковими значеннями певного типу структурної центральності, їх можна впорядкувати за значеннями обраного типу функціональної центральності і навпаки. Ці підходи також можна використати для побудови сценаріїв одночасних групових атак на систему. Так, якщо на початку впорядкованого списку є група однаково важливих за певною центральністю вузлів, то можна здійснити одночасну атаку саме на цю групу елементів МС.

Поряд із визначенням першочергових цілей ураження, не менш важливим є пошук та нейтралізація джерел такого ураження, адже часом набагато простіше усунути джерело негативного впливу на систему, ніж долати наслідки глобальної цілеспрямованої атаки чи нецільового ураження МС. Часто таке джерело є очевидним (діючий вулкан, відома терористична або хакерська група, район хімічного або радіаційного забруднення, країна-агресор тощо). У багатьох випадках його знайти достатньо просто. Так, централізовані генератори фейків характеризуються великими обсягами вихідних та мінімальними об'ємами вхідних потоків, що і дозволяє достатньо просто їх ідентифікувати за допомогою аналізу значень параметрів вхідного та вихідного впливу вузлів мережі [17]. Набагато складніша ситуація складається у випадку нецентралізованих уражень системи. Так, у більшості навіть найрозвиненіших країн світу не було виявлено так званого «нульового пацієнта», з якого почалося поширення у цих країнах Covid-19. Одним із способів нейтралізації джерела негативного впливу може бути зустрічна атака на нього, як це здійснюється під час введення економічних санкцій проти країн, що несуть загрозу світовій безпеці. Очевидно, що всі розглянуті вище підходи до захисту системи найкраще діють у поєднанні.

Вузли транспортної мережі, які забезпечують пересування найбільших об'ємів потоків в системі, та мають найбільші значення параметрів впливу та посередництва в процесі внутрішньо та міжсистемних взаємодій, підлягають першочерговому захисту від цілеспрямованих атак. Водночас, під час нецільового ураження на зразок поширення епідемій небезпечних інфекційних захворювань такі вузли потребують якнайшвидшого блокування руху пасажиропотоків. Подібні ситуації виникають під час забезпечення захисту від кібератак та поширення комп'ютерних вірусів або фейків. Тобто, блокування певної складової МС може бути як метою атаки на систему, так і способом її захисту. Звідси слідує, що проблему уразливості системи можна умовно поділити на дві задачі – пряму та обернену. Пряма задача полягає у визначенні тих елементів системи, які необхідно насамперед захистити, щоб запобігти дестабілізації роботи або припинення функціонування системи, а обернена задача зводиться до визначення тих елементів, блокування яких призведе до мінімізації втрат, які очікують систему унаслідок ураження. Як показує приклад транспортної системи, сценарії, спрямовані на підготовку захисту МС від цілеспрямованих атак, можуть бути не менш успішно використані для протидії поширенню нецільових уражень. З іншого боку, процес поширення нецільових уражень може стати основою для побудови ефективних сценаріїв цілеспрямованих атак. Так, вжиті для боротьби із поширенням пандемії Covid-19 заходи фактично перетворили світ у мережу ізольованих зон-країн, рух потоків між якими (особливо людських) із-за припинення або суттєвого обмеження залізничного, авіаційного та автомобільного сполучення зменшився на порядки. Більш того, багато держав, зокрема Україна, унаслідок введення подібних обмежень також перетворилися на мережі ізольованих зон – регіонів, громад та окремих населених пунктів. Самоізоляція більшості громадян, яка була викликана обмеженням руху транспорту, великими штрафами за недотримання умов карантину та припиненням роботи підприємств або їх роботою у режимі віддаленого доступу суттєво скоротила обсяги не лише зовнішніх, але й внутрішніх потоків у таких ізольованих зонах. За незмінної структури мережі відбулася свого роду «грануляція» системи, яка поділилася на ієрархію послідовно ізольованих у сенсі обмеження взаємодій підсистем. Ці обставини природно призвели до скорочення виробництва та торгівлі, втрати від яких суттєво пришвидшили та поглибили чергову фінансово-економічну кризу та навіть призвели до соціальних збурень в окремих країнах світу. Така «грануляція», яка стала дієвим засобом боротьби з поширенням пандемії Covid-19, водночас є дуже ефективним способом глобальної цілеспрямованої атаки на систему. Прикладом такої атаки є фінансово-економічна криза, яка почалася в Радянському Союзі унаслідок ембарго із-за війни в Афганістані та неочікувано для самих ініціаторів санкцій стала чи не най головнішим чинником розпаду СРСР на окремі незалежні республіки.

**ПОДОЛАННЯ НАСЛІДКІВ СИСТЕМНИХ УРАЖЕНЬ**

Жодна велика складна система не в змозі захистити усі свої елементи від глобальних цілеспрямованих атак або нецільових уражень різної природи. Це означає, що не менш важливим поряд із організацією захисту є розроблення засобів, спрямованих на подолання наслідків системних уражень. У статті [41] обґрунтовано необхідність створення інформаційних моделей (ІМ) важливих для життєдіяльності суспільства реальних МС для неперервного контролю за їх станом, процесом функціонування та підвищенням ефективності роботи. Не менш важливою є роль таких інформаційних моделей під час подолання наслідків негативних внутрішніх або зовнішніх впливів на МС або БШМС. Аналіз таких наслідків потребує цілісного та повного уявлення про систему як до, так і під час та після ураження. Це уявлення формується на підставі всієї інформації про історію, поточний стан та побудований на їх основі прогноз поведінки системи. Виходячи із цих міркувань, інформаційною моделлю БШМС називатимемо тотожну її структурі динамічну в сенсі постійного розширення та поповнення структуру даних, кожна компонента якої містить інформацію про стан, процес функціонування та взаємодії відповідної складової системи у поточний момент, минулому та майбутньому, починаючи з елементів та закінчуючи системою загалом. Очевидно, що ІМ реальних систем можуть містити дані надзвичайно великих обсягів, які нереально опрацювати «вручну» у прийнятні проміжки часу. Водночас, оперативний аналіз інформації та прийняття на його підставі обґрунтованих рішень є особливо важливим у кризових для системи ситуаціях [46, 47]. Одним із способів вирішення цієї проблеми є формування на основі ІМ моделі комплексного оцінювання БШМС [32, 35], до складу якої входять моделі інтерактивного та регулярного оцінювання стану та процесу її функціонування, а також так звана модель «пошуку новизни».

Модель інтерактивного оцінювання БШМС будується на підставі результатів неперервного моніторингу процесу її функціонування, які містятьсянадходять в інформаційну модель, та полягає у постійному відстежуванні взаємодії мережевих та міжшарових потоків з елементами БШМС [44]. Так, під час аналізу поширення пандемії Covid-19 завданнями цієї моделі є відбір даних про кількість нововиявлених інфікованих, людей, що одужали та померли, ко-роткострокове прогнозування необхідної кількості ліків, медичного обладнання та ліжко-місць, прийняття рішень дощо посилення або послаблення карантинних заходів і т. ін. Модель регулярного оцінювання будується на підставі інформації, зібраної за певний період часу функціонування БШМС, та передбачає глибокий та ретельний аналіз наслідків ураження усіх елементів системи, які спостерігались протягом цього періоду [32]. Регулярне оцінювання наслідків поширення Covid-19 доцільно здійснювати після завершення чергової хвилі пандемії. Завданням цієї моделі є підведення підсумків готовності державних структур та фінансово-економічної системи країни та окремих її регіонів до загроз, які виникли під час минулої хвилі, та підготовка до мінімізації негативних наслідків наступної хвилі. Модель пошуку новизни використовує всі зібрані протягом достатньо тривалого проміжку часу дані, які містяться в інформаційній моделі, та призначена для виявлення та аналізу регулярних та / або масових ефектів, нетипових за тими чи іншими ознаками для поведінки елементів системи. Основним завданням цієї моделі є виявлення загроз до їх масового поширення та перетворення в реальне ураження системи. Модель пошуку новизни доцільно використовувати для встановлення реальної картини захворюваності та смертності від Covid-19. Критерієм пошуку у даному випадку вважається позитивний результат ПЛР- або ІФА-тестування. Однак, оскільки близько 80% інфікованих переносять це захворювання без жодних симптомів і не звертаються до лікарів та далеко не всі громадяни проходять тестування, то реальну картину поширення коронавірусу натепер не встановлено. Тобто, незважаючи на наявність чітких критеріїв пошуку, сформувати об'єктивний висновок про реальний стан системи можливо лише на підставі опосередкованих даних. Важливим завданням цієї моделі також є визначення можливих віддалених наслідків цього ураження (ускладнень, дії вакцинації, погіршення стану людей з хронічними захворюваннями тощо).

У кожній із перерахованих вище моделей оцінювання використовуються адаптовані для реалізації визначених цілей моделі взаємопов'язані методи локального аналізу стану, якості функціонування та взаємодії елементів БШМС; методи агрегованого оцінювання, спрямовані на побудову узагальнених висновків стосовно окремих підсистем або шарів БШМС; та методи прогностичного оцінювання поведінки оцінок складових БШМС на коротко-, середньо- та довгострокову перспективу. Зокрема, якщо в моделях інтерактивного та регулярного оцінювання ці методи визначають негативний вплив ураження на підставі кількісної міри виходу поведінки елементів за межі встановлених стандартів, то в моделі пошуку новизни – невиявлені під час неперервного моніторингу або планових досліджень елементів системи ефекти, масово розподілені у просторі та / або часі.

Основною перевагою моделей оцінювання порівняно з ІМ БШМС є на порядки менші обсяги даних, які значно легше піддаються аналізу та дозволяють оперативно локалізувати найбільш уражені елементи системи, тобто є дієвим засобом для подолання проблеми кількісної складності системних досліджень [41]. Водночас, тотожність структур моделей оцінювання та ІМ БШМС дозволяє достатньо просто перейти від оцінки елемента до аналізу усіх наявних даних про нього, які містяться у відповідній складовій інформаційної моделі. Оскільки більшість реально існуючих БШМС є комплексними, тобто багатоцільовими та багатофункціональними утвореннями, то й їх ураження та наслідки цих уражень можуть мати комплексний характер. Саме модель комплексного оцінювання БШМС, у якій поєднуються різні види системних досліджень, методи інтелектуального опрацювання даних та взаємопов'язані підходи до аналізу результатів дозволяє створювати адекватну картину поведінки системи та приймати відповідні оперативні рішення по усуненню виявлених загроз.

## ВИСНОВКИ

У 2020-2022 роках людство зіштовхнулося з двома глобальними викликами, перший з яких (пандемія Covid-19) є яскравим прикладом загальносистемного нецільового ураження, а другий (російсько-українська війна) – цілеспрямованої атаки (нападу рф на Україну) та викликаної нею загрози світової продовольчої, енергетичної, фінансової кризи і зворотні всеосяжні санкції стосовно агресора, негативні наслідки яких торкнулися практично всіх країн світу. Людство виявилося непідготовленим до таких викликів, але на часі залишаються проблеми глобального потепління, кліматичних катастроф та масштабних посух. За минуле півстоліття спостерігаються безпрецедентно швидкі темпи втрати біорозмаїття у живій природі [48], а протягом останніх 20 років витрати на боротьбу з кліматичними катастрофами зросли у 8 разів [49]. Натепер ученим відомо більш ніж 20 вірусів небезпечних інфекційних захворювань, мутації яких можуть призвести до поширення пандемій, значно катастрофічніших за Covid-19 [50], посилюється загроза глобальних військових конфліктів тощо. Це підтверджує актуальність вивчення особливостей групових та загальносистемних уражень різних видів та розроблення методів захисту від них для багатьох реальних систем. Нецільові негативні впливи або цілеспрямовані атаки на систему чи міжсистемні взаємодії можуть бути спрямованими на дестабілізацію їх структури та/або процесу функціонування, виникати унаслідок дії як внутрішніх, так і зовнішніх джерел. Розуміння структурної та функціональної важливості елементів системи дає змогу обирати об'єкти, які потребують першочергового захисту або найбільш сприяють поширенню ураження. Розроблення сценаріїв захисту або блокування таких об'єктів та практична реалізація цих сценаріїв, своєчасна нейтралізація джерел або причин ураження дозволяє значно зменшити шкоду, яку можуть нанести ці ураження як окремим МС, так і БШМС, до складу яких вони входять. Комплексне оцінювання наслідків негативного впливу на систему або процес міжсистемних взаємодій та визначення послідовно-сті дій, за допомогою яких можна подолати ці наслідки та повернути їх до неураженого стану є не менш важливим кроком для забезпечення нормальної життєдіяльності.

**ПЕРЕЛІК ПОСИЛАНЬ**